\documentclass[reprint,superscriptaddress,nofootinbib,amsmath,amssymb,aps]{revtex4-1}
\usepackage{graphicx}
\usepackage{xcolor}
\usepackage{dcolumn}
\usepackage{bm}
\usepackage{braket}
\usepackage{mathrsfs}
\usepackage{hyperref}
\hypersetup{
    colorlinks,%
    citecolor=blue,%
    filecolor=blue,%
    linkcolor=blue,%
   	urlcolor=blue,
   	linktoc=page
}

\def\bz{{\bar z}}
\def\bw{{\bar w}}
\def\bzeta{{\bar\zeta}}
\def\p{\partial }
\newcommand{\N}{\mathscr{N}}
\newcommand{\C}{\mathscr{C}}
\newcommand{\T}{\mathcal{T}}
\renewcommand{\P}{\mathcal{P}}
\newcommand{\D}{\mathcal{D}}
\newcommand{\V}{\mathbb{V}}
\newcommand{\M}{\mathbb{M}}
 \newcommand{\badat}{\begin{alignedat}}
 \newcommand{\eadat}{\end{alignedat}}

\begin{document}

\title{BMS particles}

\author{Xavier Bekaert}
 \email{xavier.bekaert@univ-tours.fr }
\affiliation{%
Institut Denis Poisson, CNRS-UMR 7013, Université de Tours, 
Parc de Grandmont, 37200 Tours, France 
}
\author{Laura Donnay}
 \email{ldonnay@sissa.it}
 \affiliation{%
 {{SISSA,
Via Bonomea 265, 34136 Trieste, Italy}}\\
{{INFN, Sezione di Trieste,
Via Valerio 2, 34127, Italy}}
}%
\author{Yannick Herfray}
 \email{yannick.herfray@univ-tours.fr }
\affiliation{%
Institut Denis Poisson, CNRS-UMR 7013, Université de Tours, 
Parc de Grandmont, 37200 Tours, France 
}


\begin{abstract}
We construct wavefunctions for unitary irreducible representations (UIRs) of the Bondi-Metzner-Sachs (BMS) group, i.e. BMS particles, and show that they describe quantum superpositions of (Poincaré) particles propagating on inequivalent gravity vacua. This follows from reconsidering McCarthy's classification of BMS group UIRs through a unique, Lorentz-invariant but non-linear, decomposition of supermomenta into hard and soft pieces.
\end{abstract}

\maketitle

\section{Introduction}
\label{sec:Introduction}

While the BMS group appeared long ago as the asymptotic symmetry group of flat spacetimes~\cite{Bondi:1962px,sachs_gravitational_1961,Sachs:1962zza}, the fact that these symmetries act simultaneously on past and future null infinity was only understood recently~\cite{Strominger:2013jfa}. The global nature of BMS symmetry  was critical in Strominger's demonstration of the BMS-invariance of the gravitational $S$-matrix \cite{Strominger:2013jfa,He:2014laa,Strominger:2017zoo}. 

Previous disregard for these asymptotic symmetries is deeply related to the longstanding problem \cite{Mott_1931,Sommerfeld:1931qaf,Bloch:1937pw} that scattering amplitudes involving massless particles are plagued with infrared (IR) divergences, rendering the $S$-matrix ill-defined. Indeed, as understood in \cite{Kapec:2017tkm} for electrodynamics (see also \cite{Gabai:2016kuf}) and later extended to the gravitational case in \cite{Choi:2017ylo}, IR divergences arise,  for conventional states, as a necessity 
to respect BMS conservation laws.

\indent In the AdS/CFT holographic correspondence, both bulk states and boundary operators organize into unitary irreducible representations (UIR)
of the conformal group. If there exists a holographic dual theory to quantum gravity in asymptotically flat spacetime, the correspondence between boundary and bulk states should similarly be given by an equivalence of UIR representations of the BMS group (see~\cite{Arcioni:2003td,Arcioni:2003xx,Dappiaggi:2004kv,Dappiaggi:2005ci} for early works).

In a series of pioneering works, McCarthy studied and classified $\textrm{BMS}_4$ UIRs \cite{Mccarthy:1972ry,McCarthy_72-I,McCarthy_73-II,McCarthy_73-III,McCarthy_75,McCarthy_76-IV,McCarthy_78,McCarthy_78errata} (see also \cite{Cantoni:1966,Cantoni:1967,Girardello:1974sq} and, for BMS$_3$ UIRs, \cite{Barnich:2014kra,Barnich:2015uva,Oblak:2016eij}). While McCarthy's results provide a foundational basis, from a physics perspective they raise just as many questions as they answer. Notably, there are infinitely more BMS representations than usual Poincaré particles. Out of these, only the representations identified by Sachs~\cite{sachs_asymptotic_1962} (for massless particles) and Longhi--Materassi~\cite{Longhi:1997zt} (for massive particles) are well understood. This is because these very specific BMS particles, referred to as `hard' in the following, are in 1:1 correspondence with usual Poincaré particles. However, we are still left with infinitely many BMS particles that require a physical interpretation. The above discussion on IR divergences suggests that these particles should correspond to some form of IR-modified states. The aim of this Letter is to provide such realizations, based on group-theoretical considerations.

Infrared gravitons are associated with two related features  \cite{Ashtekar:1987tt,Strominger:2013jfa}: the possibility of carrying `soft' BMS charges (or `memory' charges) and the existence of `boundary gravitons' (or `supertranslation Goldstone' modes). The possibility of having nonzero soft charge $\partial_z^2\partial_{\bz}^2\N(z,\bz)$ is what makes BMS particles infinitely richer than their Poincaré counterparts. At the level of representation theory this translates \cite{Bekaert:2025kjb} into the fact that the supermomentum $\P(z,\bz)$ can be uniquely decomposed as 
\begin{equation}\label{Supermomentum decomposition1}
   \P(z,\bz) = \partial_{z}^2\partial_{\bz}^2\N(z,\bz) + P(z,\bz)\,,
\end{equation}
where the hard contribution $P(z,\bz)$ is a (non-linear) function of the momentum $p_{\mu}$ only: $P(z,\bz)=\omega \delta^{(2)}(z-\zeta)$ if $p_{\mu}=\omega q_{\mu}(\zeta,\bzeta)$ is null, $P(z,\bz)= -\frac{m^4}{\pi}\big(p\cdot q(z,\bz)\big)^{-3}$ if $p^2 = -m^2 \neq0$ (see below for our conventions). Boundary gravitons $\partial_z^2\C(z,\bz)$ parametrize \emph{gravity vacua}, i.e. possible -- diffeomorphic but inequivalent -- backgrounds. As we shall demonstrate by constructing the corresponding wavefunctions, BMS particles are best thought of as quantum superpositions of Poincaré particles propagating on different gravity vacua.

The Letter is organized as follows. We start with a review of the gravitational phase space in
Section \ref{sec:phase_space}, where we introduce mode expansions for the asymptotic radiative data as well as the soft graviton mode and supertranslation Goldstone mode. The interpretation of the latter as the mode labeling different gravity vacua is illustrated by the Kirchhoff-d’Adhémar formula in Section \ref{sec:vacua}. BMS wavefunctions are presented in Section \ref{sec:BMS_wavefunction} and related to BMS UIRs in Section \ref{sec:BMS_particles}. The section \ref{sec:first_quant} closes the loop by explaining how these states are realized in terms of Strominger's phase space. Finally, in Section \ref{sec:Discussion}, we summarize the results and discuss some implications of this Letter.

\section{Gravitational phase space}
\label{sec:phase_space}
The mode expansion of a graviton of momentum $p^\mu=(\,p^0,\vec p\,)$ is given by
\begin{equation}\label{Gravitational phase space: linearized spin 2}
    \hat{h}_{\mu\nu}(X) = \kappa \int \!\!\frac{d^3 p}{(2\pi)^3\, 2p^0} \Big[\varepsilon_{\mu\nu}^{*\alpha}(\vec p)\, \hat a_\alpha(\vec p) \, e^{i p\cdot X} + \text{h.c.}\Big]\,,
\end{equation}
where $X^{\mu} \in \mathbb{R}^{3,1}$, $\kappa=\sqrt{32\pi G}$ and $\alpha=\pm$ are the two helicities. Using the parametrization $p^\mu=\omega q^\mu(z,\bz)$, with the null vector $q^\mu=\frac{1}{\sqrt 2}\big(1+z\bz, z+\bz,-i(z-\bz),1-z\bz\big)$, the polarization tensors can be written as $\varepsilon^{\pm\mu \nu}=\varepsilon^{\pm\mu}\varepsilon^{\pm\nu}$ with $\varepsilon^{+\mu}=\p_z q^\mu$, $\varepsilon^{-\mu}=\p_\bz q^\mu$; see e.g. \cite{He:2014laa}. The commutation relations for annihilation/creation operators, $\hat a_{\alpha}(\omega)$, $\hat a_{\alpha}^{\dagger}(\omega)$ ($\omega \neq0$), 
are then given by
\begin{equation}\label{Commutation relations, hard}
\!\big[\hat a_{\alpha}(\omega,z), \hat a_{\alpha'}^{\dagger}(\omega',z')\big]  \!=\! \frac{2 (2\pi)^3}{\omega} \delta(\omega-\omega')\delta(z-z')\delta_{\alpha\alpha'}.
\end{equation}
In flat Bondi coordinates, where the Minkowski metric is $ds^2=-2 du dr+2r^2 dz d\bz
$, the large $r$ expansion of the graviton mode yields
 $dx^{\mu}dx^{\nu}\hat{h}_{\mu\nu}(X) \sim r\, \hat C_{zz}(u,z,\bar z) dz^2  + \text{h.c.} + \mathcal O(r^0)$, where the asymptotic shear $\hat C_{zz}$ is given by~\cite{He:2014laa} 
\begin{align}\label{Gravitational phase space: shear}
    \!\!\!\hat C_{zz} \!=\! \frac{ \kappa }{8i \pi^2}\!\!\int_0^{\infty}\!\!\!\! d \omega \Big( \hat a_+(\omega, z,\bar z) e^{- i \omega u}\!- \hat{a}^{\dagger}_-(\omega, z,\bar z) e^{i \omega u}\!\Big).
\end{align}
A similar expression holds for $\hat C_{\bz\bz}$, which encodes the other graviton helicity.
As shown in \cite{Strominger:2013jfa,He:2014laa}, the gravitational phase space at null infinity must include, in addition to the radiative modes $\hat a_\alpha(\omega)$, $\hat a^\dagger_\alpha(\omega)$ ($\omega \neq 0$), the soft mode $\hat{\mathscr N}(z,\bz)$ and its symplectic partner, the Goldstone boson $\hat \C(z,\bz)$ of spontaneously broken supertranslation
invariance. They can be related to the late and early time behavior of the shear as 
\begin{align*}
-\partial_z^2 \hat{\mathscr N}(z,\bar z)& =\tfrac{4}{\kappa^2} \lim_{u\to \infty}\big( \hat C_{zz}(u,z,\bar z) - \hat C_{zz}(-u,z,\bar z) \big)\,,\\[0.1em]
-2\partial_z^2 \hat{\mathscr C}(z,\bar z)& = \tfrac{1}{2} \lim_{u\to \infty}\big( \hat C_{zz}(u,z,\bar z) +  \hat C_{zz}(-u,z,\bar z) \big)\,.
\end{align*}
As emphasized by Ashtekar, e.g. in \cite{Ashtekar:1987tt,Ashtekar:2014zsa}, the presence of these modes implies that Sachs' norm for the shear is not finite. How, then, should we think of quantum states in the presence of soft charge? This is one of the questions we shall answer.

The soft and Goldstone modes commute with $\hat a_{\alpha}(\omega)$, $\hat a^{\dagger}_{\alpha}(\omega)$ ($\omega \neq0$) and the only non-vanishing commutator is~\cite{He:2014laa}
\begin{align}\label{Commutation relations, soft}
[\partial^2_{\bar z}\hat{ \mathscr N}(z,\bar{z}),  \partial^2_w \hat{ \mathscr C}(w,\bar w)]  = -i\, \delta^{(2)}(z-w)\,.
\end{align}

Under the action of the group $\textrm{BMS}_4 \simeq SL(2,\mathbb{C}) \ltimes C^{\infty}(S^2)$, the hard operators transform as
\begin{equation}
 \!\! \!\!\hat a_\pm'(\omega',z',\bar z') \!=\! \left(\frac{\partial z'}{\partial z}\right)^{\!\!\mp 1}\!\!\!\left(\frac{\partial \bar z'}{\partial \bar z}\right)^{\!\!\pm 1}\!\!\!\!\!e^{-i\omega \mathcal{T}(z,\bz)}\,\hat a_\pm(\omega,z,\bar z)\,, \label{transfo a poincare}
\end{equation}
where $z' =\frac{az +b}{cz+d}$ and $\T(z,\bz)\in C^{\infty}(S^2)$ are the corresponding Möbius transformation and supertranslations, respectively. 
Soft operators both transform as $SL(2,\mathbb C)$ primaries of weights $(-\tfrac{1}{2},-\tfrac{1}{2})$,
\begin{equation}\label{Gravitational phase space: BMS action on G and N}
    \badat{2}
     \! \!\! \hat \N'(z',\bz')&=\left(\frac{\partial z'}{\partial z}\right)^{-\frac{1}{2}}\!\!\!\left(\frac{\partial \bar z'}{\partial \bar z}\right)^{-\frac{1}{2}}\! \hat \N(z,\bz)\,,\\
 \! \!\! \hat \C'(z',\bz')&=\left(\frac{\partial z'}{\partial z}\right)^{-\frac{1}{2}}\!\!\!\left(\frac{\partial \bar z'}{\partial \bar z}\right)^{-\frac{1}{2}} \!\Big[\hat \C(z,\bz)+\T(z,\bz)\Big]\,,
    \eadat
\end{equation}
while the action of supertranslations induces a \emph{shift} of the Goldstone mode. As a result, Minkowski spacetime transforms under BMS supertranslations as
\begin{equation}
    \badat{2}
      \!\!  \left(ds^2\right)' =& -2 du dr+2r^2 dz d\bz \\ & -2 r\left( \partial^2_{z} \T dz^2 + \, \partial^2_{\bz} \T d\bz^2 \right) + \mathcal O(r^{0})\,.
    \eadat
\end{equation}
The space of all possible Minkowski spacetimes obtained in this way will be denoted $\V$  and will be referred to as the space of gravity vacua.

\section{Gravity vacua}
\label{sec:vacua}

The mode $\partial^2_{z} \C$, labeling the different gravity vacua, and the radiative data $\hat{a}_{\alpha}(\omega)$ ($\omega \neq0$) play a very different role in the theory: the field $\partial^2_{z} \C$  defines the background, denoted $\mathbb{M}_{\C}$, on which the radiative data propagate.

Following Newman \cite{Newman:1976gc}, the field $\partial^2_{z} \C$ can indeed be understood as the piece of data needed to reconstruct Minkowski spacetime in a holographic manner from null infinity $\mathscr{I}$. The key idea here is to consider the space $\Gamma[\mathscr{I}]$ of cuts of null infinity i.e. the space  of sections
\begin{equation}
   U :\left| \begin{array}{ccc}
        S^2 &  \to & \mathscr{I} \\
        (z,\bz) & \mapsto &\big(u =U(z,\bz), z, \bz\big).
    \end{array}\right.
\end{equation}
There are infinitely many of these sections $U\in\Gamma[\mathscr{I}]$. Picking a shear that is pure gauge $C_{\bz\bz} = -2\partial_{\bz}^2 \C $, one can however consistently \emph{define} Minkowski spacetime $\mathbb{M}_{\C} \subset \Gamma[\mathscr{I}]$ as the subspace of real solutions to the `good cut' equation \cite{Footnote1}
\begin{align}
 U &\in \mathbb{M}_{\C}    & \Leftrightarrow&& \partial_{\bz}^2 U &= \frac{1}{2} C_{\bz\bz}.
\end{align}
A generic solution must be indeed of the form
\begin{equation}\label{goodcutsol}
    U_X(z,\bz) = -\C(z,\bz) - q_{\mu}(z,\bz) X^{\mu}.
\end{equation}
It is however clear that, through this procedure, one can construct infinitely many \emph{different} copies $\mathbb{M}_{\C}\in \mathbb{V}$ of Minkowski spacetime, altogether forming the space of gravity vacua $\V$. Each of these gravity vacua can be related to the reference flat spacetime $\mathbb{M}_{\C=0}$ via a supertranslation.

Once a background $\mathbb{M}_{\C}$ has been chosen, one can propagate the radiative data. This is best illustrated by the Kirchhoff-d'Adh\'emar formula, cf. \cite{Penrose1980GoldenON} or \cite[Chap. 5.12]{Penrose:1985bww}, which allows one to reconstruct the bulk field $h_{\mu\nu} (X)$ from the joint data of the boundary value $C_{zz}$ at $\mathscr{I}$ and the gravity vacuum $\mathbb{M}_{\C}$: 
\begin{equation}\label{KA}
    \!\!\!\!h_{\mu\nu} (X,\C) \!=\!  -\frac{1}{2\pi}  \hspace{-0,4cm}  \underset{  \phantom{ph}u= U_X}{\int} \hspace{-0,4cm} d^2 z \;\varepsilon^{*+}_{\mu\nu}(z,\bar z)  \partial_{u}C_{zz}(u,z,\bz)  + \text{h.c.}
\end{equation} 
Equivalently,  making use of \eqref{Gravitational phase space: shear} and \eqref{goodcutsol},
\begin{align}\label{hard BMS wavefunction1}
    \!\!&h_{\mu\nu} (X, \C)\\
    &= \frac{\kappa}{16\pi^3}\! \int \!\omega d\omega d^2 z \Big[\varepsilon_{\mu\nu}^{*\alpha}(z,\bar z)\, a_\alpha(\omega, z,\bar z)  e^{i \omega (q\cdot X+\C)}  + \text{h.c.} \Big],\nonumber
\end{align}
which, when $\C=0$ and upon quantization, is seen to give back formula \eqref{Gravitational phase space: linearized spin 2}. Note that going from \eqref{Gravitational phase space: linearized spin 2} to \eqref{hard BMS wavefunction1} amounts to replacing the creation/annihilation operators by the dressed operators of \cite{Himwich:2020rro}. The above is the wavefunction of a hard (massless) BMS particle. This is the simplest instance of a BMS particle. 

\section{BMS wavefunctions}
\label{sec:BMS_wavefunction}

Equation \eqref{hard BMS wavefunction1} suggests to write a general (massless) BMS wavefunction as
\begin{equation}\label{BMS wavefunction in BMS space}
 \!\! |\Psi \rangle  = \int_{\mathbb{V}}\! \D \C \int_{\mathbb{M}_{\C}} \!\!\! d^4X  \;\; \Psi\left( X  ; \partial^2\C \right) | X ; \partial^2 \C \rangle\,,
\end{equation}
with $\partial_{\mu}\partial^{\mu}\Psi=0$ at fixed $\C$. Here $| X ; \partial^2 \C \rangle$ is an eigenstate of both position and gravity vacuum
\begin{align*}
    \hat{X}^{\mu}| X ; \partial^2 \C \rangle &\!=\! X^{\mu} | X ; \partial^2 \C \rangle,& \!\!\!\!  \partial_z^2 \hat{\C} | X ; \partial^2 \C \rangle \!=\! \partial_z^2 \C | X ; \partial^2 \C \rangle
\end{align*}
and the first integral is a path integral over the space of gravity vacua. The above form will ultimately be justified by the fact that all BMS UIRs can be realized in this way, see section \ref{sec:BMS_particles}. We will also later show how to relate it to Strominger's gravitational phase space \eqref{Commutation relations, hard}-\eqref{Commutation relations, soft}. Let us here only highlight some essential features.

Keeping $\C$ fixed gives the wavefunction of a field in a given gravity vacuum $\M_{\C}$,
\begin{equation}\label{wavefctinagivenvacuum}
\int_{\mathbb{M}_{\C}} \!\!\! d^4X  \;\; \Psi\left( X  ; \partial^2\C \right) | X ; \partial^2 \C \rangle\,.
\end{equation}
Therefore, the generic state \eqref{BMS wavefunction in BMS space} is a quantum superposition of such wavefunctions in all possible gravity vacua $\M_{\C}\in\V$.

The `Fourier transform' in $X$, 
\begin{align}\label{eigenstates: hybrid eigenstate from position eigenstate}
   \!\!\! | \omega, z, \bz; \partial^2 \C \rangle \!&:=\!\! \int_{\M_{\C}} \!\!\!\!\!\!d^4X\, e^{i\omega \big(q(z,\bz)\,\cdot X +\C(z,\bz)\big)}| X ; \partial^2 \C \rangle,
\end{align}
yields eigenstates of momentum in a given gravity vacuum, and \eqref{BMS wavefunction in BMS space} can then be rewritten as
\begin{equation}
 \!\! |\Psi \rangle  \!=\! \int_{\V} \! \D \C \!\!\int\! \omega d \omega d^2 z \! \;  \Psi\left( \omega, z, \bz ; \partial^2\C \right) | \omega, z, \bz; \partial^2 \C \rangle
\end{equation}
where
\begin{align*}
   \Psi\!\!\left( X, \partial^2\C \right) \!&=\!\!  \int \!\omega d\omega d^2 z\,  e^{i \omega \big(q(z,\bz)\cdot X+\C(z,\bz)\big)}\Psi\!\!\left( \omega, z, \bz ; \partial^2\C \right).
\end{align*}
In the hard case \eqref{hard BMS wavefunction1}, this last equation gives
\begin{align}\label{hard BMS wavefunction2}
\Psi^{hard}\left( \omega, z, \bz ; \partial^2\C \right)=
\begin{cases}
     a^{\dagger}(\omega, z,\bz) & \omega>0\\
   a(\omega, z,\bz) &\omega<0
\end{cases}.
\end{align}

\section{BMS particles}
\label{sec:BMS_particles}
In usual Poincaré representation theory (\`a la Wigner), a particle is given by a function $\Psi(P)$ in momentum space $\left(\mathbb{R}^{3,1}\right)^*$, defined as the dual space  to the space of translations. More precisely, Poincar\'e particles (i.e. UIRs of the Poincar\'e group)  are described by wavefunctions that only have support on a given orbit of the Lorentz group (e.g. the null cone for massless particles), the spin being given by a choice of UIR for the little group $\ell_P$\,.

A BMS particle \cite{Mccarthy:1972ry,McCarthy_78,McCarthy_78errata,McCarthy_75,McCarthy_73-II,McCarthy_72-I,McCarthy_73-III,McCarthy_76-IV} is described by a wavefunction $\Psi(\P)$ in supermomentum space (i.e. $(C^{\infty}(S^2))^*$, the dual space to the space of supertranslations):
\begin{equation}\label{BMS particles: wavefunction in supermomentum space}
    |\Psi\rangle = \int \D \P\; \Psi(\P)\; | \P \rangle.
\end{equation}
As the notation suggests, $|\P\rangle$ is here an eigenstate of the supermomentum operator $\hat{\P}(z,\bz)$, which is the generator of supertranslations. Since supertranslations have weights $(-\frac{1}{2}, -\frac{1}{2})$, supermomenta must have weights $(\frac{3}{2}, \frac{3}{2})$, the duality pairing being given by
\begin{equation}
    \langle \P , \T \rangle = \int d^2z \,\P(z,\bz) \,\T(z,\bz)\,.
\end{equation}
The space of states of the form \eqref{BMS particles: wavefunction in supermomentum space} always forms a BMS representation: an element of the BMS group $(M,\T) \in SL(2,\mathbb{C}) \ltimes C^{\infty}(S^2) $ acts as
\begin{align}\label{BMS particles: rep of BMS in supermomentum space}
   |\Psi\rangle \mapsto \int \D \P \;e^{i \langle \P, \T \rangle}\;\Psi(\P\cdot M)\; | \P \rangle\,.
\end{align}
where $\P\cdot M$ indicates that Möbius transformations act from the right on supermomenta. 
The UIRs of the BMS group are then given by functions which only have support on a given orbit $\mathcal{O}_{\P}$ of the Lorentz group. These orbits generalize the mass-shell of usual Poincaré particles; the BMS equivalent of spin being given by a choice of UIR for the BMS little group $\ell_{\P} \subset SL(2,\mathbb{C})$, defined as the stabilizer of a supermomenta $\P\in \mathcal{O}_{\P}$ in the orbit. The heart of the seminal work of McCarthy \cite{Mccarthy:1972ry,McCarthy_78,McCarthy_78errata,McCarthy_75,McCarthy_73-II,McCarthy_72-I,McCarthy_73-III,McCarthy_76-IV} was to classify all possible little groups appearing in this way.

The existence of a Lorentz-invariant projection from supermomenta $\P$ to momenta $p_{\mu}$,
\begin{equation}\label{BMS particles: projection from super to momenta}
    \P(z,\bz) \mapsto \pi_{\mu}(\P)  =\int \!d^2z\, q_{\mu}(z,\bz) \,\P(z,\bz)
\end{equation}
means that one can talk about the Poincaré little group $\ell_{\pi(\P)}$ of a BMS particle and, in particular, define its mass square $m^2= \pi_\mu(\P)\,\pi^\mu(\P)$. It follows from \eqref{BMS particles: projection from super to momenta} that the BMS little group is necessarily contained in the Poincaré little group
\begin{equation}
    \ell_{\P} \subseteq \ell_{\pi(\P)} \subseteq SL(2,\mathbb{C})\,.
\end{equation}
In this Letter, we will only discuss massless BMS representations, i.e.
\begin{equation}\label{BMS particls: momentum}
    \pi_{\mu}(\P) = \omega\, q_{\mu}(\zeta,\bzeta)\,.  
\end{equation}
See \cite{Bekaert:2025kjb} for a discussion on the massive case (which does not alter significantly the picture described here). Of crucial importance is the possibility to always decompose a massless supermomentum
as
\begin{equation}\label{Supermomentum decomposition}
    \P(z,\bz) = \partial_z^2 \partial_{\bz}^2 \N(z,\bz) + \omega\,\delta^{(2)}(z-\zeta)\,.
\end{equation}
We wish to emphasize that this decomposition is unique and Lorentz-invariant. However, it is \emph{not} stable under addition, due to the nonlinear dependence in the momentum \eqref{BMS particls: momentum} of the hard contribution $P(z,\bz)= \omega\, \delta^{(2)}(z-\zeta)$. The decomposition \eqref{Supermomentum decomposition} means that, introducing the notation $| \omega, \zeta,\bzeta; \partial^2\! \N\rangle$ for the eigenstates of $\hat{\P}$ of eigenvalue \eqref{Supermomentum decomposition}, one can always write a massless BMS particle as
\begin{equation}\label{BMS particles: wavefunction in supermomentum space2}
   \!\!\! |\Psi\rangle \! = \!\!\int \!\!\omega d \omega d^2 \zeta \!\!\int\!\! \D \!\N \,\Psi(\omega, \zeta,\bzeta; \partial^2\! \N)| \omega, \zeta,\bzeta; \partial^2 \!\N\rangle,
\end{equation}
where the wavefunction $\Psi(\omega, \zeta,\bzeta; \partial^2 \!\N)$ must only have support on a given orbit $\mathcal{O}_{\P}$ of the Lorentz group inside the space of supermomenta. To be more explicit, factorize elements of the Lorentz group as
\begin{align}
      M(p, W
      ) = L(p)\, W
      \in SL(2,\mathbb{C}),
\end{align}
where $p\mapsto L(p)$ is an injective map from the null cone to $SL(2,\mathbb{C})$ and $W\in \ell_p \subset SL(2,\mathbb{C})$. Let $\N^0(z,\bz)$ be a reference soft charge defining the orbit of the BMS particle and define $\N_p := \N^0\cdot L(p)^{-1}$. One can then write \cite{Footnote3}
\begin{align}
\label{BMS particles: wavefunction in supermomentum space3}
    \!\!\! |\Psi\rangle \! = \!\!\int \!\!d^3 p \!\!\int_{\ell_{p} / \ell_{\P}}\hspace{-0,7cm} dW
    \;\Psi\big(p; (\partial^2 \N_p) \cdot W
    \big)\; | p; (\partial^2 \N_p)\cdot W
    \rangle\,.
\end{align}
More generally, BMS particles are always realized as \emph{finite-dimensional integrals} of dimension
\begin{equation}\label{Dimension of the orbit}
    \dim(\mathcal{O}_{\P}) = 6 - \dim(\ell_{\P}).
\end{equation}
As explained in \cite{Bekaert:2025kjb}, for hard representations $\dim(\ell_{\P}) = \dim(\ell_{\pi(\P)}) = 3$ and one recovers the usual  mass-shell dimension 3. However, a generic orbit will in fact have no non-trivial stabilizer and thus will have support on a six-dimensional submanifold in the space of supermomenta. All intermediate dimensions are also allowed and correspond to non-trivial BMS little groups (see e.g. \cite{McCarthy_75,McCarthy_76-IV,McCarthy_78,McCarthy_78errata}).

Since supermomenta $\P$ are dual to supertranslations $\T$ one can define the (infinite-dimensional) Fourier transform
\begin{equation}
    | \T \rangle = \int \D \P\, e^{-i\,\langle \P, \T \rangle} | \P \rangle
\end{equation}
and rewrite BMS particles as 
\begin{equation}\label{BMS particles: wavefunction in BMS space}
  \!\!  |\Psi\rangle = \int \D \T\, \Psi(\T) \;| \T \rangle.
\end{equation}
Once again, while the wavefunction in BMS space $\Psi(\T)$ appears as a complicated functional on the (infinite-dimensional)  space of supertranslations, let us emphasize that, once Fourier transformed to supermomentum space, the wavefunction $\Psi(\P)$ of a BMS particle only lives in a submanifold of dimension \eqref{Dimension of the orbit}. Finally, since a supertranslation $\T$ can always be decomposed as
\begin{equation}\label{supertranslation hard/soft decomposition}
    \T(z,\bz) = \C(z,\bz) + q_\mu(z,\bz) X^\mu
\end{equation}
where $\C = \T\big|_{l\geqslant 2}$ only contains higher spherical harmonics of $\T$ and $q\cdot X = \T\big|_{l=0,1}$ the lower ones, one can write $\Psi( X, \partial^2\C) := \Psi(\T)$ and recover the expression \eqref{BMS wavefunction in BMS space} for a BMS state.
Note that, while the decomposition \eqref{supertranslation hard/soft decomposition} is not Lorentz-invariant, the projection $\T \mapsto \partial_{z}^2 \C$
is Lorentz-invariant. Thus, it makes sense to talk of the restriction of a BMS wavefunction \eqref{BMS wavefunction in BMS space} to a fixed gravity vacuum $\partial_z^2\C$ as in \eqref{wavefctinagivenvacuum}. Similarly, one can restrict a BMS UIR to the subspace of wavefunctions in a fixed gravity vacuum. Such a subspace is not BMS-invariant, but it is invariant under the corresponding Poincar\'e group.

To conclude the comparison with Section \ref{sec:vacua}, we note that eigenstates of momentum in a given gravity vacuum \eqref{eigenstates: hybrid eigenstate from position eigenstate} are related to supermomentum eigenstates as
\begin{align}\label{eigenstates: supermomentum eigenstate from hybrid eigenstate}
   \!\! | \omega, z, \bz; \partial^2\! \N \rangle \!&=\!\! \int \!\D\C  e^{i\int d^2w\,\partial_w^2 \N \partial_{\bw}^2 \C} |  \omega, z, \bz ;  \partial^2 \C \rangle.
\end{align}

\section{First quantization of gravity vacua}
\label{sec:first_quant}

In this final section, we relate the BMS particle states \eqref{BMS wavefunction in BMS space} to Strominger's gravitational phase space \eqref{Commutation relations, hard}-\eqref{Commutation relations, soft}. Let us first recall that the supermomentum operator is realized as~\cite{He:2014laa}
\begin{align}\label{Supermomentum operator expression}
   \!\!\hat{\mathcal P}(z,\bz)  
   &=\p_\bz^2\p_z^2 \hat \N \!+\frac{1}{16\pi^3}\!\!\int_0^\infty \!\!\!\!d\omega\, \omega^2 (\hat a_+^\dagger\hat a_++\hat a_-^\dagger\hat a_-).
\end{align}
The above split between soft and hard parts was also studied in \cite{Donnay:2021wrk} where it was checked that each piece transforms separately in the coadjoint representation of the BMS algebra \cite{Barnich:2021dta}. 

Now, the fact that the Goldstone current is shifted under the action of supertranslations \eqref{Gravitational phase space: BMS action on G and N} suggests to think of $\partial_z^2 \hat{\C}(z,\bz)$ as a \emph{position operator} in the space of gravity vacua $\mathbb{V}$. The commutator \eqref{Commutation relations, soft} is then seen to be the canonical commutation relation $[\hat{x},\hat{p}]= i\hbar$ in the space of gravity vacua. It follows from the commutation relations \eqref{Commutation relations, hard}-\eqref{Commutation relations, soft} and the expression \eqref{Supermomentum operator expression} for the supermomentum operator that supermomentum eigenstates of eigenvalue \eqref{Supermomentum decomposition}, $|\P\rangle = | \omega, \zeta, \bzeta ; \partial^2\! \N \rangle$, are obtained as \cite{Footnote2}
\begin{align}
    | \omega, \zeta, \bzeta ; \partial^2 \!\N \rangle &=\, e^{i\,\langle \partial^2 \bar{\partial}^2 \!\N , \hat{\C} \rangle}\,\hat a^{\dagger}(\omega, \zeta,\bzeta)  |0\rangle,
\end{align}
where $|0\rangle$  denotes the BMS vacuum 
which, by definition, satisfies
\begin{align}
    \hat a_{\alpha}(\omega, z,\bz)|0\rangle=&\,0,&  \partial_z^2\hat{\N}(z,\bz)|0\rangle=&\,0.
\end{align}
BMS wavefunctions \eqref{BMS wavefunction in BMS space} are then constructed from \eqref{BMS particles: wavefunction in supermomentum space3} and the successive Fourier transforms \eqref{eigenstates: hybrid eigenstate from position eigenstate} and \eqref{eigenstates: supermomentum eigenstate from hybrid eigenstate}. As we saw, they should be thought of as quantum superpositions of usual Poincaré particles in different gravity vacua. Among all possible states that can be constructed in this way, states of the form 
\begin{align}
    \!\!\! | p \rangle_{\C} \! = \!\int_{\ell_{p} / \ell_{\P}}\hspace{-0,7cm} dL
    \,\;\;e^{i\,\langle\,(\partial^2 \bar{\partial}^2 \N)\cdot L
    ,\, \C\,\rangle}\; | p; (\partial^2\! \N)\cdot L\rangle\,
\end{align}
stand out as BMS particles of momentum $p_{\mu}$ and `localized' in a gravity vacuum $\partial_z^2 \C$. As opposed to the eigenstates \eqref{eigenstates: hybrid eigenstate from position eigenstate} of  $\partial^2_z\hat{\C}$, the above states belong to a UIR of BMS and are thus genuine particles. Finally, note that for hard particles, $\N=0$, $\ell_\P = \ell_p$ and the $\C$ dependence drops out so that, by construction, such representations cannot discriminate a particular vacuum.

\section{Discussion}
\label{sec:Discussion}

In the present Letter, we showed that BMS particles, i.e. BMS UIRs, can always be realized in terms of wavefunctions of the form \eqref{BMS particles: wavefunction in BMS space}. These can naturally be interpreted as quantum superpositions of usual particles, each of them propagating on a different gravity vacuum. The (familiar) hard representation \eqref{hard BMS wavefunction1} stands out by propagating the same particle in all possible gravity vacua, see \eqref{hard BMS wavefunction2}. The essence of our construction was to reconsider McCarthy's results in light of the hard/soft decomposition of supermomenta \eqref{Supermomentum decomposition1} suggested by recent developments \cite{Strominger:2017zoo}; see \cite{Bekaert:2025kjb} for more details.  Given the close similarity of asymptotic symmetries~\cite{Footnote4}, the present construction carries through almost identically for QED. 

The physical picture emerging from the present Letter fits very naturally with Refs. \cite{Kapec:2021eug,Kapec:2022axw,Kapec:2022hih} suggesting to think of $S$-matrix observables as living ``over the moduli space of vacua''. It is also in line with the reinterpretation \cite{Kapec:2017tkm} of Faddeev-Kulish (FK) states \cite{Kulish:1970ut}, which led to the generalized states of \cite{Choi:2017ylo}: as already emphasized in the massive case in \cite{Chatterjee:2017zeb}, hard massless particles, whose supermomenta are of the form $\omega \delta^{(2)}(z-\zeta)$, are not enough to ensure conservation of supermomentum. In the language of \cite{Kapec:2017tkm}, this means that hard states cannot ensure the conservation of BMS charges, 
which is the reason for IR divergences. Reference \cite{Kulish:1970ut} then amounts to constructing states of supermomenta $\omega \,\delta^{(2)}(z-\zeta) - \frac{\omega}{2\pi}\, \partial_{\bar z}^2\big(\frac{\bar z-\bzeta}{z-\zeta}\big)$. Keeping in mind that these are weighted distributions, these supermomenta do not vanish and one can show that, for such states, conservation of momentum implies conservation of supermomentum \cite{SM}. From this perspective, the FK construction is rather unnatural and it is simpler to just require the total conservation of BMS charges in order to obtain IR-finite $S$-matrix elements \cite{Kapec:2017tkm,Choi:2017ylo}. 

It follows that if an IR-finite unitary $S$-matrix for massless particles can be defined, then it will have to be BMS-invariant. Therefore the asymptotic one-particle states must belong to UIRs of the BMS group and the multi-particle states must be suitable tensor products thereof. In the present Letter, we gave a physical realization of the corresponding one-particle states and showed that Strominger's phase space can be understood as a quantization scheme where gravity vacua are first-quantized while the remaining hard degrees of freedom are second-quantized. It should be clear from our presentation that a complete first quantization can be obtained by considering functions on the (infinite-dimensional) homogeneous space BMS$_4/SO(3,1)$. However, it is also clear that one should rather consider a complete second quantization as the starting point for an extension of the massless $S$-matrix. This is where, we believe, our work relates and contrasts with the recent works \cite{Prabhu:2022zcr,Prabhu:2024zwl,Prabhu:2024lmg}, which consider a Hilbert space including supermomentum eigenstates. 
It appears that the latter does not coincide with the Fock space of BMS particles. The perspective put forward in this Letter is rather to stick to Fock quantization of BMS particles, i.e. asymptotic multi-particle states constructed as tensor products $|\omega_1, \zeta_1, \bar{\zeta}_1;\partial^2 \!\N_1\rangle\otimes\cdots\otimes |\omega_n, \zeta_n,\bzeta_n;\partial^2 \!\N_n\rangle$ of free one-particle states spanning BMS UIRs. S-matrix elements of such BMS particles are structurally different from usual ones, due to the extra degrees of freedom involved. As an initial step in constructing such amplitudes, and investigating their potential to resolve infrared divergences, one should identify the equivalent of Mandelstam variables in this context. Additionally, one should develop appropriate limiting procedures (see \cite{Bekaert:2025kjb} for early suggestions) to establish connections with the conventional S-matrix and observable cross-sections. We will return to these questions in upcoming works. 

\begin{acknowledgments}
The authors thank Sasha Zhiboedov and Hofie Hannesdottir for discussions.
L.D. is supported by the European Research Council (ERC) Project 101076737 -- CeleBH. Views and opinions expressed are however those of the author only and do not necessarily reflect those of the European Union or the European Research Council. Neither the European Union nor the granting authority can be held responsible for them.
L.D. is also partially supported by INFN Iniziativa Specifica ST\&FI.  Her research was also supported in part by the Simons Foundation through the Simons Foundation Emmy Noether Fellows Program at Perimeter Institute. Research at Perimeter Institute is supported in part by the Government of Canada through the Department of Innovation, Science and Economic Development and by the Province of Ontario through the Ministry of Colleges and Universities. L.D. and Y.H. thank IDP and SISSA for the hospitality during their respective visits.
\end{acknowledgments}

\bibliographystyle{style}
\bibliography{references}

\begin{thebibliography}{10}
\bibitem{Footnote1}
\textnormal{From the perspective of Newman's H-space theory, we are here only dealing with the trivial case where spacetime is flat. In particular, the solutions are real.}
\bibitem{Footnote2}
\textnormal{Notice that this expression should not be understood as the action of a supermomentum creation operator and that the vacuum used here should not be confused with the zero-particle state of a Fock representation.}
\bibitem{Footnote4}
The asymptotic group of QED is the semi-direct product $SO(3,1) \ltimes ( \mathbb{R}^{3,1} \times \mathcal{E}[0])$.
\bibitem{Footnote5}
Notice that in this reference the convention for the delta fonction is $\int i dz \wedge d\bz\, \delta^{(2)}(z)=2$, as opposed to the present article where we take $\int i dz \wedge d\bz\, \delta^{(2)}(z)=1$.
\end{thebibliography}

\clearpage
\appendix
\addcontentsline{toc}{section}{Supplemental Material}
\onecolumngrid
\begin{center}
   \large\textbf{\MakeUppercase{Supplemental Material}} \\[0.8em]  
    
    
    \large \textbf{``BMS particles''} \\[0.2em]
    
    \normalsize Xavier Bekaert, Laura Donnay, Yannick Herfray\\[2.4em]
\end{center}
\twocolumngrid
\section*{Comparison with (revisited) Faddeev-Kulish states}
In this supplemental material, we contrast our BMS particle states with the revisited Faddeev-Kulish states first considered in \cite{Kapec:2017tkm} for electrodynamics and later adapted in the gravitational case in \cite{Choi:2017ylo}. The state
\begin{align}
     |\psi\rangle=&\, e^{\hat{R}}\,\hat a^{\dagger}(\omega, \zeta,\bzeta)  |0\rangle,
\end{align}
involves the dressing operator 
\begin{align}\label{R_factor}
  \hat{R}&= \frac{\kappa}{2}\int \frac{d^3k}{16\pi^3 k^0} \,f^{\mu\nu}(k)\left[ \hat a^{\dagger}_{\mu\nu}(k) - \hat a_{\mu\nu}(k)\right]\\
  &= \frac{\kappa}{2}\int \frac{\varpi d\varpi d^2z}{16\pi^3} \left( f_+(\hat a^{\dagger}_- - \hat a_+) + f_-(\hat a^{\dagger}_+ - \hat a_-)  \right)\,,\nonumber
\end{align}
with $k^{\mu}= \varpi q^{\mu}(z,\bz)$ and $f_{\mu\nu} = \sum_{\alpha} f_{\alpha} \epsilon^{\alpha}_{\mu\nu}$ a real symmetric tensor, where the sum runs over polarizations. The state $|\psi\rangle$ is an eigenstate of the soft part of the supermomentum operator defined in Eq. (34) in the main text, $\hat{\P}^{soft}(z,\bz) = \partial_z^2 \partial_{\bz}^2\hat{\N}(z,\bz)$, namely
\begin{equation}
  \hat{\P}^{soft}(z,\bz) |\psi\rangle =   \partial_{\bz}^2\partial_{z}^2\N(z,\bz) |\psi\rangle\,,
\end{equation}
where $\partial_z^2\partial_{\bz}^2 \N = \frac{1}{4\pi} \lim\limits_{\varpi\to 0}\varpi\Big(\partial_{z}^2 f_+  + \partial_{\bz}^2 f_-\Big)$. 
It is however not eigenstate of the full supermomentum operator $\hat{\P}(z,\bz)$ (given in Eq. (34) in the main text), due to the fact that the operator \eqref{R_factor} inserts infinitely many hard particles.

Let us now compare this with an eigenstate $|\mathcal P\rangle$ of the full supermomentum operator, given in Eq. (35) in the main text. From the above and the definition of the pairing, one can write
\begin{align*}
   i \langle \partial_z^2 \partial_{\bz}^2 \N \!, \hat{\C} \rangle\! =\!i  \int \frac{\varpi d\varpi d^2z}{2\pi} \left( f_+ \delta(\varpi)\partial_z^2 \hat{\C} + f_- \delta(\varpi)\partial_{\bz}^2 \hat{\C} \right).
\end{align*}
Comparing with \eqref{R_factor}, one sees that the Faddeev-Kulish dressing consists in adding to the supermomentum eigenstate infinitely many hard particles.

The original Faddeev-Kulish amplitudes \cite{Kulish:1970ut} can be obtained \cite{Kapec:2017tkm,Choi:2017ylo} by taking $f^{\mu\nu} = \frac{p^{\mu}p^{\nu}}{p\cdot k} \, \psi(p,k)$ where $p^{\mu} = \omega q^{\mu}(\zeta,\bar{\zeta})$, with the condition that $\psi(p,k)=1$ close to $\omega=0$. Therefore
\begin{equation}\label{f FK}
    f_- = -\frac{\omega}{\varpi} \frac{\bz-\bar{\zeta}}{z-\zeta}\, \psi(p,k)\,,
\end{equation}
and $f_+ = (f_-)^*$, which implies that the state $|\psi\rangle$ is an eigenstate of $\hat{\P}^{soft}(z,\bz)$ of eigenvalue
\begin{equation}\label{FK eigenvalue}
    \partial_{\bz}^2\partial_{z}^2\N = -\frac{\omega}{2\pi}  \partial_{\bz}^2 \left(\frac{\bz-\bar{\zeta}}{z-\zeta}\right)\,.
\end{equation}
It might be worth stressing that this last statement is, as always when supermomenta are involved, a statement about distributions, i.e.
\begin{equation}
    \langle \hat{\P}^{soft} , \T \rangle |\psi\rangle  = \left(-\frac{\omega}{2\pi}\int \!d^2z\, \partial_{\bz}^2\T(z,\bz) \,\frac{\bz-\bar{\zeta}}{z-\zeta} \right)\, |\psi\rangle.
\end{equation}
With the choice \eqref{FK eigenvalue}, the state
\begin{align}\label{à la FK}
    | \omega, \zeta, \bzeta ; \partial^2 \!\N \rangle &=\, e^{i\,\langle \partial^2 \bar{\partial}^2 \!\N , \hat{\C} \rangle}\,\hat a^{\dagger}(\omega, \zeta,\bzeta)  |0\rangle\,,
\end{align}
is an eigenstate of supermomentum $\hat{\P}(z,\bz)$ of eigenvalue
\begin{equation}
    \omega \delta^{(2)}(z-\zeta) - \frac{\omega}{2\pi} \partial_{\bz}^2 \left(\frac{\bz-\bar{\zeta}}{z-\zeta}\right) = p^{\mu}(\zeta,\bar{\zeta})\mathcal{D}_{\mu}\delta^{(2)}(z-\infty)\,,
\end{equation}
where the right-hand side is a distribution supported at $z=\infty$. The proof of this last identity, as well as the expression for $\mathcal{D}_{\mu}\delta(z-\infty)$, can be found in \cite{Bekaert:2025kjb} (see Proposition 3.4) \cite{Footnote5}. 

Therefore, one sees that the dressing \emph{à la} Faddeev-Kulish \eqref{à la FK} together with \eqref{FK eigenvalue}, takes a state of supermomentum $\omega \delta^{(2)}(z-\zeta)$ to a state with (non-zero) supermomentum $p^{\mu}\mathcal{D}_{\mu}\delta(z-\infty)$. The effect is thus to make the supermomentum of the dressed state \emph{linear} in the momentum, trivially ensuring conservation of supermomentum as a result of the conservation of momentum. Notice that the price to pay for this is that the dressing breaks Lorentz invariance, as manifested by the special role that the point $z=\infty$ acquires; in the usual Faddeev-Kulish approach, this comes from the fact that the choice leading to expression \eqref{f FK} for $f_- = f^{\mu\nu}\epsilon_{\mu\nu}^+$ is not gauge invariant since $f^{\mu\nu}k_{\nu}\neq0$.

\end{document}